\documentclass{aa}   
\usepackage{graphicx}
\usepackage{natbib}
\usepackage{txfonts}
\usepackage{amssymb}
\usepackage[T1]{fontenc}
\newcommand{\ii}{\'{\i}}

\newcommand{\cd}{cd$^{-1}$}

\newcommand{\msun}{M$_{\sun}$}

\newcommand{\kms}{$\mathrm{km}$\,$\mathrm{s}^{-1}$}

\begin{document}

   \title{ HD~172189: An eclipsing and spectroscopic binary with a
   $\delta$~Sct-type pulsating component in an open cluster}
   \titlerunning{HD~172189: An eclipsing and spectroscopic binary with a
     $\delta$~Sct pulsating component in an open cluster}
   
   \author{S. Mart\'{\i}n-Ruiz\inst{1} \and P.\,J. Amado\inst{1,2} \and
   J.\,C. Su\'arez\inst{1} \and A. Moya\inst{3} \and A. Arellano
   Ferro\inst{4} \and I. Ribas\inst{5} \and E. Poretti\inst{6}}

   \offprints{S. Mart\'{\i}n-Ruiz}

   \institute{Instituto de Astrof\'{\i}sica de Andaluc\'{\i}a, CSIC, apdo.
              3004, E-18080 Granada, Spain
         \and European Southern Observatory, Alonso de Cordova 3107,
              Santiago 19, Chile 
         \and LESIA, Observatoire de Paris-Meudon, 92195 Meudon Principal
              Cedex, France
         \and Instituto de Astronom\ii a, Universidad Nacional Aut\'onoma de
              Mexico, Apdo. Postal 70-264, 04510 Mexico D.F., Mexico
         \and Institut d'Estudis Espacials de Catalunya/CSIC, Campus UAB,
              Facultat de Ci\`encies, Torre C5-parell-2a planta, 08193
              Bellaterra, Spain 
         \and INAF-Osservatorio Astronomico di Brera, Via Bianchi 46, 23807 
              Merate (LC), Italy}

   \date{Received ... / Accepted ...}

   \abstract{We present a study based on more than 2500 $uvby$--measurements
     collected on HD\,172189, a new 
     eclipsing binary system with a $\delta$ Sct-type pulsating component 
     that belongs to the open cluster IC\,4756. The great interest of this
     object lies in that three important characteristics coexist: cluster
     membership, binarity and pulsation. Its binarity and the $\delta$
     Sct-type pulsations of one component were detected in the course of
     several Str\"omgren photometric campaigns. The frequency analysis of all
     the $uvby$ out-of-eclipse data reveals a clear frequency of 19.5974 \cd as
     well as the existence of other high values in the range 18-20 \cd. 
     We have also
     carried out a first evaluation of the binary parameters and determined a
     value of 5.702~d for the orbital period. An additional potential interest
     of HD\,172189 resides in its location in the field of view of the COROT
     mission and, therefore, this star is an excellent candidate for becoming 
     a target for asteroseismology.
  
            }

\maketitle


\section{Introduction\label{sec:intro}}

The existence of stellar pulsation in an eclipsing binary system is a very
interesting phenomenon for the theoretical study of the variable component
because binarity provides useful information on its stellar parameters and
pulsation modes \citep{Reed02}. However, few objects of this type have
been discovered to date. Several examples of $\delta$ Sct pulsators in
stellar systems can be found in \citet{Lampens00}. More recently,
\citet{Rodriguez04} report a complete list of $\delta$ Sct stars
detected in eclipsing binary systems. In addition, if binarity and
intrinsic variability are favourable conditions that contribute to a
better knowledge of the pulsational behaviour, the fact that the star
belongs to an open cluster implies additional advantages. The stellar
parameters (age, chemical composition and distance), which are shared by
all the member stars of a cluster, place constraints on stellar models,
enhancing our knowledge from the oscillation modes. RS Cha is the only
example of an eclipsing binary with a $\delta$ Sct component in a cluster
has been reported in the literature. This star is an A-type double-lined
eclipsing binary belonging to the new young open cluster $\eta$
Chamaeleontis \citep[] [and references therein]{Luhman04}. Long before its
membership to this cluster was established, RS Cha was discovered to be an
eclipsing binary by \citet{Strohmeier64}. A $uvby$ study to determine the
parameters of the system was carried out by \citet{Clausen80} but the
authors were not able to conclude whether the detected $\delta$ Sct
variation correspond to only one or both components. Recently,
\citet{Gaspar03} have discovered a $\delta$ Sct star in an eclipsing 
binary in the field of the cluster NGC\,2126 using CCD
photometric observations. However, their astrometric study suggests
non-membership to the cluster.
     
HD\,172189 is classified as an A2-type star with a $V$ magnitude of 8.85.
Several studies using proper motions demonstrate that this object belongs
to the open cluster IC\,4756 (star 93 in the numbering system of
\citep{Kopff43}). The binarity of HD\,172189 was discovered in the summer
of 1997 over the course of an observing campaign carried out for detecting
$\gamma$ Doradus pulsating stars in the cluster IC\,4756
\citep{Martin00th,Martin03p}. With a time span of 10 nights, only one
minimum of light was detected. In order to determine the orbital period of
the binary system, new observing campaigns were carried out in following
years. From these measurements, new light minima and $\delta$ Sct-type
oscillations were detected. 

Furthermore, HD\,172189 is located in the 
summer field of view of the forthcoming COROT
mission \citep{Baglin02}. Lying close to the main target HD\,171384, it
represents an excellent candidate for being selected as secondary target
\citep{Poretti05}.

\section{Observations\label{sec:obs}}


The $uvby$ photometric observations presented in this letter were
obtained in different campaigns during 1997, 2003 and 2004, lasting for a
total of 48 nights. In the years 1997 and 2003, the $uvby$ measurements
were performed using the 90-cm telescope at Sierra Nevada Observatory
(SNO, Spain) and later, during the summer of 2004, a two-site campaign was
carried out together with the 1.5-m telescope at San Pedro M\'artir
Observatory (SPMO, Mexico). In both sites, twin six-channel simultaneous
Str\"omgren photometers were used. We employed HD\,172365, SAO\,123720,
HD\,173369, HD\,181414 as comparison and check stars for the different
observing runs. In the 2004 coordinated campaign, only one comparison star
(HD\,172365) was observed because the results derived from the first
observations did not show any sign of variability. In total, more than
2500 measurements in each of the four filters $uvby$ have been obtained.

Moreover, two closely time spaced, high--resolution ($R$=48000) spectra 
were collected in June 2, 2004 using the FEROS spectrograph attached to 
the ESO 2.2 m-telescope at La Silla (Chile). The spectra were obtained 
in the framework of the preparation of the GAUDI archive \citep{Solano05} 
for the COROT mission. 

\section{The membership of HD\,172189 to the cluster IC\,4756 \label{sec:ic4756}}

The first reference to the open cluster \object{IC\,4756} dates from 1943,
when \citet{Kopff43} carried out a detailed study of the clusters
\object{NGC\,6633} and IC\,4756 and provided spectral types for 203 stars
using photographic photometry and spectroscopy. Several photometric
studies have been published with estimations of the cluster's age,
distance or colour excess \citep{Alcaino65, Schmidt78, Smith83,
Martin00th}. However, the number of studies on the membership of the
observed stars is smaller.  Preliminary proper motions and $UBV$
photometry were obtained by \citet{Seggewiss68}, whose values were updated
by \citet{Herzog75}. In this latter work, the membership of HD\,172189 to
the open cluster IC\,4756 was negative according to the $UBV$ photometry.
Only those stars located within twice the standard error of the
photographic photometry to the ZAMS and giant branches in the two-colour
and the colour-magnitude diagrams simultaneously were considered as
members. \citet{Schmidt78} later obtained Str\"omgren photometric
measurements of HD\,172189 ($b-y = 0.258$, $m_1 = 0.123$, $c_1 = 1.055$,
$\beta = 2.820$) and derived a colour excess of $E(b-y)=0.159$ and a true
distance modulus of 7$^m$11. This distance modulus value is considerably
smaller than the cluster average 8$^m$05 in spite of the reddening value
being similar to what was expected for the corresponding part of the
cluster where HD\,172189 is located ($\sim$0.15 mag). The conclusions of
both investigations about the membership of HD\,172189 to the cluster
based on photometry should thus be revised. The effects of the binarity we
claim may modify such conclusions since apparent magnitudes as well as
colour indices were considered as representing those of a single star.

The opposite situation is found when considering the analysis of the
membership from proper motion measurements. Firstly, \citet{Herzog75}
reported a probability of 0.92 of HD\,172189 belonging to IC\,4756. More
recently, the membership of this star was confirmed by \citet{Missana95}
who re-studied this cluster through a specific statistical method. Their
results, obtained with the $\chi^2$ criterion and those obtained by
\citet{Herzog75} from the criterion of the maximum likelihood of
\citet{Sanders71} are in concordance, reporting for HD\,172189 a
probability of 0.95 for cluster membership.

\section{The binary system}

\begin{figure}
\includegraphics[width=5cm,angle=-90]{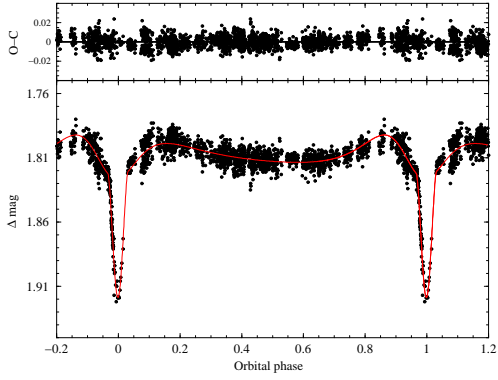}
   \caption{Str\"omgren differential $v$ light curve of HD~172189 for all 
data with a preliminary fit (see text). The upper panel shows the (O-C) 
residuals.}
\label{fig:lcs}
\end{figure}

The light curves from the data collected for HD~172189 show three light
minima due to eclipses. Only one of the eclipses was followed past the
minimum in brightness. We carried out a period search using the method of
\citet{Schwarzenberg96}. From the analysis we obtained the ephemeris $T
\mbox{(min I)} = \mbox{HJD}2452914.644(3) + 5.70198(4)$ d. The Str\"omgren
$v$-filter differential magnitude phased with the orbital period of
5.70198~d is depicted in the bottom panel of Fig.~\ref{fig:lcs}.  The
observed primary eclipse depth is of approximately 0.12 mag.

We used the May 2003 version of the Wilson-Devinney program
\citep{Wilson71} for a first modelling of the photometric light curve. We
fitted the orbital parameters to reproduce the depth of the primary
eclipse, the lack of a secondary minimum and the out-of-eclipse
variations. The solutions call for a system with a moderate orbital
eccentricity of $e\approx 0.24$, a longitude of the periastron of $\omega
\approx 68^{\circ}$ and an inclination of $i\approx 73^{\circ}$.  The best
fit is obtained with two components of quite different radii ($r_2/r_1
\approx 0.6$) but similar temperatures (${T_{\rm eff}}_2/{T_{\rm eff}}_1
\approx 1.05$), although the temperature ratio is a poorly constrained
parameter.  It is worth noting that the absence of a secondary minimum is
likely caused by the geometric orbital configuration (i.e., different
relative separations at the two eclipse phases) rather than by the
components having very unequal temperatures. This lack of a secondary
eclipse makes the determination of $e$ and $\omega$ still tentative
because these parameters are fully determined by the subtle out-of-eclipse
variations. In Fig.~\ref{fig:lcs} we show our best fit to the Str\"omgren
$v$ light curve, with an rms of 0.0057 mag, and the phased (O-C)
residuals.

The available high-resolution spectra were used to measure the radial
velocities of the two components since double lines are clearly visible
(SB2). We employed the two-dimensional cross-correlation algorithm
$\mathrm{TODCOR}$ \citep{Zucker94} with synthetic spectrum templates
computed for a temperature of 8000~K. We obtained velocities of about
63~\kms and $-$127~\kms for the primary (i.e., more luminous) and
secondary components, respectively. Unfortunately, the two spectra were
taken very close in time and in the same orbital quadrature and thus the
direct determination of the component mass ratio is not possible.
Nevertheless, a rough estimation of the mass ratio can be made if we
assume the mean radial velocity of the cluster ($\sim$$-$26~\kms;
\citealt{Mermilliod90}) as the systemic velocity of the binary. We obtain
a value of $M_2/M_1\approx 0.9$, which suggests that the two components
are quite similar in mass. In addition, our analysis yields a luminosity 
ratio of the two components of $L_2/L_1\approx 0.5$, which, when 
combined with the mass ratio, would indicate a rather evolved system. 
According to the ephemeris listed above, the spectra were taken around 
phase 0.83. However, because of the still uncertain values of $e$ and 
$\omega$, we cannot make a reliable prediction of the individual stellar 
masses at this point. A complete radial velocity curve, as well as 
multi-wavelength and dense light curves, are needed to achieve a full 
description of the binary system.

Finally, a possible alternative explanation for the observation of only
one eclipse in the light curve is the presence of identical eclipses
separated by 0.5 phase (i.e., zero eccentricity). In this case, the
orbital period of the binary system would be about 11.4 d. Note, however,
that this scenario results in a light curve with out-of-eclipse variations
that cannot be easily explained with fitting programs. Moreover, from the
measured radial velocities, $K_1+K_2\ga\mid RV_1-RV_2\mid = 190$~\kms can 
be estimated, with $K_1$ and $K_2$ being the velocity semi-amplitudes. 
Using Kepler's Third law, the total mass of the system
in this case would be greater than 8~\msun, predicting components with
masses significantly higher than expected for a $\delta$ Sct star. 
We thus strongly favour the scenario with $P=5.702$~d.

\section{The pulsating component}

\begin{figure}
    \includegraphics[width=8.6cm]{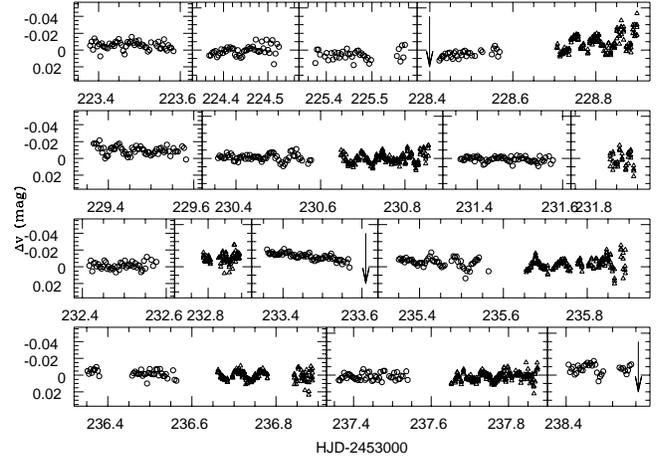}
   \caption{$v$-band differential light curves obtained from the two-site 
     campaign. The primary eclipse phases have not been plotted. The open 
     circles and triangles
     correspond to the SNO and SPMO data, respectively. The vertical 
     arrows mark the positions of the three observed primary minima 
     (HJD-2453000 = 228.174, 233.875 and 239.578).}
   \label{fig:lcurve}
\end{figure}

The short period variability caused by pulsations during
out-of-primary-eclipse observations is shown in
Fig.~\ref{fig:lcurve}. This
differential light curve in the $v$ filter corresponds to the last
two-site campaign in the summer of 2004. Very small-amplitude oscillations
around a few thousandths of a magnitude are observed in most of the
nights. The deviations with respect to the average magnitude are produced
by the variations of the orbital phase of the binary pair. As can be seen, 
this magnitude difference becomes more evident
near the primary eclipses, for instance, around HJD 2453228.4-2453228.8
and HJD 2453233.4.

As to the identity of the pulsating component, there is no reliable
evidence at this point. However, the oscillations are probably
related to the primary component since it contributes $\sim$70\% of the
total light. To determine the oscillation modes of the pulsation a
frequency analysis has been performed using all out-of-primary-eclipse
measurements. The method used is the same as in \citet{Rodriguez98}, where
single frequency and multiple frequency techniques are combined using both
Fourier and multiple least squares algorithms.

\begin{figure}
\begin{center}
  \includegraphics[width=7.4cm]{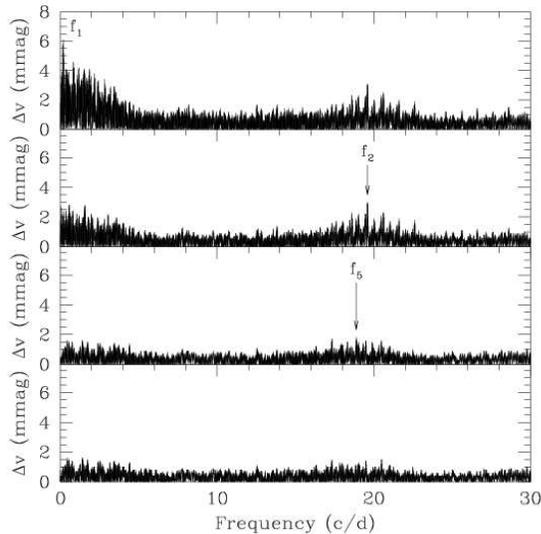}
   \caption{Power spectra in the $v$ filter before and after prewhitening 
   the following frequences: $f_1 = 0.17541$ \cd, $f_2= 19.59740$ \cd and 
  $f_5 = 18.87939$ \cd.}
  \label{fig:espectro}
\end{center}
\end{figure}

To ensure consistent results, the analysis has been made in the three
filters ($vby$) where the precision of the measurements is better. In
addition to the frequency $f_1=0.17541$ \cd corresponding to the orbital
period, we have detected a set of peaks in the typical $\delta$ Sct
region of 18-19 \cd, with $f_2=19.59740$ \cd and $f_5=18.87939$ \cd
corresponding to the frequencies with the highest amplitudes.
Figure~\ref{fig:espectro} shows the power spectra in the filter $v$ before
(top panel) and after prewhitening the orbital period, $f_2$ and $f_5$
(middle and bottom panels), respectively. The bottom panel represents the
residuals. After this cleaning procedure, several peaks remain in high
frequencies with semi-amplitudes below 2 mmag.

Our analysis not only shows peaks corresponding to the high modes but also
reveals the presence of low frequencies -- $f_3 = 1.55425$ \cd and $f_4 =
0.05995$ \cd -- not shown in Fig.~\ref{fig:espectro}. In contrast to $f_2$
and $f_5$, these frequencies were different in all filters. Such values
appear as a consequence of the binary effects, not completely removed by the 
$f_1$ term, together with a high number of measurements collected from
only one site and the existence of the numerous gaps between the different
observing periods. For the purpose of verifying the nature of $f_3$ and
$f_4$, a second frequency analysis has been carried out with only the data
set observed during the 2004 two-site campaign. The results confirm 
the orbital period obtained considering all the measurements in addition 
to new frequencies very close to $f_2$ (19.5960 and 19.4887 \cd). The number 
of peaks in low frequencies, as well as the number of aliases, decreases. 
Therefore, the values $f_3$ and $f_4$ have been excluded from the final 
solution. The spectral region in which peaks from the binarity variation and
data sampling appear, is sufficiently far from that of the $\delta$ Sct 
variation as to not to affect our analysis of the pulsations.

\section{Conclusions}

Photometric measurements collected over several years have allowed us to
discover the binary nature and $\delta$ Sct pulsation of HD\,172189. The
first minimum in brightness was observed during the summer of 1997. To
detect a second eclipse, new campaigns were carried out in 2003, which
also unveiled small amplitude oscillations in the light curves. The
orbital period of the binary system could not be determined until the
two-site campaign in 2004 took place. The most important preliminary
results of these observations are:
\vspace{-2mm} 
\begin{enumerate}
\item An orbital period of 5.702~d and an eclipse depth in the primary
minimum of about 0.12~mag have been estimated.
\item Preliminary simulations of the binary solution suggest that this object
is a classic eclipsing binary system with an eccentric orbit. The components
likely have similar masses and are at a rather evolved evolutionary stage.
\item Frequencies have been detected in the region of 18-20 \cd when analyzing
the out-of-eclipse data. Within this range, the frequences of 19.59740 and
18.87939 \cd are those with higher amplitudes (less than 5 mmag).
\end{enumerate}

With high probability, HD\,172189 is a binary system belonging to the
IC\,4756 open cluster that also shows $\delta$ Sct-type pulsations. We
have presented a preliminary analysis of the currently available data but
more photometric and spectroscopic observations are required to verify the
physical parameters of this eclipsing binary system. Measurements obtained
during the eclipses could be very useful in typifying the oscillation modes
of the pulsating component. Actually, HD\,172189 offers an unique possibility
in the $\delta$ Sct scenario, especially when considering the possible COROT
pointing. Indeed, the complete characterization of this important
system will permit detailed investigations of stellar structure and
evolution thanks to the many constraints provided by binarity and cluster
membership.

\begin{acknowledgements} 
Thanks are especially due to the staff of the Sierra Nevada Observatory
for assistance during the observations and Monica Rainer for the reduction and
calibration of the FEROS spectra. We thank Eloy Rodr\ii guez and Victor
Costa for their valuable suggestions. S.\,M.-R. acknowledges the financial support, as
a 'Averroes' postdoctoral contract, funded by the Junta de Andaluc\ii a. 
J.\,C.\,S. thanks the Direcci\'on General de Investigaci\'on for supporting his
work under project AYA2003-04651 and the Junta de Andaluc\ii a. A.\,A.\,F. is 
gratefull to DGAPA-UNAM project IN110102 for financial support. I.\,R. 
acknowledges support from the Spanish Ministerio de Ciencia y Tecnolog\'{\i}a
through a Ram\'on y Cajal fellowship.
\end{acknowledgements}


\bibliography{/home/users/dfe/susana/macros/styles/susana.bib}
\bibliographystyle{aa}


 \end{document}